# ELECTRIC SURFACE POTENTIAL OF DODECANE NANO-DROPS IN AQUEOUS SOLUTIONS OF LOW IONIC STRENGTH

Daniela Díaz [1], Kareem Rahn-Chique [2], German Urbina-Villalba [2,*]

[1] Instituto Universitario Tecnológico "Dr. Federico Rivero Palacio", Carretera Panamericana, Km. 8, Caracas, Venezuela.

[2] Instituto Venezolano de Investigaciones Científicas (IVIC), Centro de Estudios Interdisciplinarios de la Física (CEIF), Carretera Panamericana Km. 11, Aptdo. 20632, Caracas, Venezuela. Email: german.urbina@gmail.com

**Abstract** While the surface charge of solid particles is a direct consequence of their synthesis, the one of suspended oil drops depends on the adsorption equilibrium of the surrounding molecules. The presence of salt raises the polarity of the water phase, favoring the salting out of the surfactant from the aqueous solution and increasing its surface excess. Yet, the electrolyte also screens the resulting surface charge of the drops [Debye-Hückel, 1923]. As a result, the electrostatic surface potential increases with the ionic strength until the saturation of the interface and then decreases. This behavior produces a maximum previously observed in hexadecane-in-water nanoemulsions [Urbina-Villalba, 2013; 2015].

Here, the variation of the surface potential of two dodecane-in-water (d/w) nanoemulsions is evaluated as a function of the sodium chloride concentration. As expected, maximum values are obtained for concentrations of 0.5 and 7.5 mM sodium dodecylsulfate (SDS). However, the surface excess of dodecane drops shows an intermediate behavior between the adsorption equilibrium predicted by macroscopic adsorption isotherms, and the one previously found on hexadecane drops.

The stability of the prepared emulsions was followed monitoring the change in the average radius of the dispersions during five minutes, six times the lapse of time employed in a typical evaluation of the aggregation rate. In the case of 7.5 mM SDS, the smallest change in size coincides with the maximum surface potential found (40 mM NaCl). This is not observed for 0.5 mM SDS. Three regimes of destabilization are found to exist depending on the ionic strength of the aqueous phase. They correspond to the prevalence of: 1) solubilization/ripening, 2) aggregation, and 3) crystal precipitation.

## 1. INTRODUCTION

During the last few years, the Laboratory of Physical Chemistry of Colloids made an appreciable effort in developing theoretical tools for the prediction of emulsion stability. Such research requires a continuous contrast between experimental evidence and simulations. This implies the development of novel analytical tools for the evaluation of destabilization rates along with the usual implementation of characterization techniques. In this regard, aggregation rates of ionic nanoemulsions are now commonly evaluated in our lab for dilute [Rahn-Chique, 2012a; 2012b; 2012c] and concentrated emulsions [Rahn-Chique, 2017a]. They are deduced from the increase of the absorbance of the system as a function of time caused by a sudden injection of brine. Unfortunately, flocculation rates of dilute nanoemulsions can only be measured above 300 mM NaCl, since lower ionic strengths are insufficient to screen the electrostatic charge of the drops appreciably, not causing the desired aggregation behavior during a reasonable sampling time.

Generally, the forecast of emulsion stability requires appropriate consideration of surfactant phase behavior. Low volume fraction nanoemulsions ($\phi \approx 10^{-4}$) can be completely solubilized at "sub-cmc" concentrations (< 8.3 mM SDS) in the presence of salt [Rahn-Chique, 2017b]. This phenomenon does not occur if the surfactant concentration is too low (0.5 mM SDS). Hence, for [NaCl] < 300 mM NaCl, a 0.5 mM SDS nanoemulsion is stable, but counter-intuitively, a 7.5-mM SDS dispersion is not. The latter showing a decrease of the absorbance with time until complete solubilization of the drops [Díaz, 2018]. Thus, it is not possible to appraise the aggregation rate of these systems by the observation of the absorbance over long periods of time.

Above 300 mM NaCl, the surface charge of the drops is sufficiently screened, and flocculation occurs faster than solubilization. Hence, the absorbance of 0.5 and 7.5 mM SDS nanoemulsions increases in a similar fashion as a result of secondary minimum aggregation [Urbina-Villalba, 2015; 2016]; the slope of the absorbance depending on the amount of salt. Above 600 mM NaCl (20° < T < 25°C), a concentration of 7.5 mM SDS stimulates the formation of surfactant crystals favoring aggregation rates substantially slower than the ones of 0.5 mM SDS [Díaz, 2018]. In this case, as it happens in the absence of salt, 7.5 mM SDS emulsions appear to be more stable than their 0.5 mM SDS analogs.

It is clear from above that a simulation of emulsion stability cannot resemble the experimental behavior unless it accounts for surfactant solubilization and precipitation, besides surfactant adsorption.

Back in 2017 we implemented a simple routine to incorporate micelle solubilization in d/w and hexadecane-in-water (h/w) simulations [Rahn-Chique, 2017b]. It was found that the results depend critically on the solubilization rate ($S_R = dR/dt$) of the drops, and the maximum number of oil molecules that can be incorporated by a micelle ($N_{oil}^{max}$). These parameters were formerly approximated by the solubilization rates of the oils *in the absence of salt* (1.92 x $10^{-13}$ m/s for hexadecane, 4.49 x $10^{-11}$ m/s for dodecane [Ariyaprakai, 2008]), and the maximum micelle solubilization of hexadecane ($N_{oil}^{max} \sim N_h^{max} = 2$). However, it is expected that $S_R$ changes appreciably with the ionic strength of the solution, since it depends on the surfactant concentration: $25 < N_d^{max} < 54$ for 35 mM < [SDS] < 139 mM [Ariyaprakai, 2008]. According to our experiments, it requires between 30 and 45 minutes to completely solubilize a d/w emulsion of ϕ = 3.2 x $10^{-4}$ at 300 mM NaCl. According to our simulations, this process takes 29 minutes assuming $N_d^{max} = \infty$.

Among the limitations that hinder the ordinary use of full-scale simulations, the short time step required to sample the repulsive potential between the drops is by far the most severe (7.9 x $10^{-8}$ s). The evaluation of an aggregation rate using a 500-drop calculation requires a year of real time in a Dell Workstation with 2.4-GHz Xeon processors. This fact favored the implementation of two alternative approaches: a) 25-particle simulations to appraise the qualitative behavior of the systems, and b) two-particle simulations to evaluate stability ratios ($W_{11}$) [Urbina-Villalba, 2016]. This latter procedure allows the computation of several flocculation rates using only *one* 500-particle simulation at very high ionic strength (1 M NaCl).

Use of two-particle simulations for 7.5-mM SDS h/w emulsions showed that an ad hoc "nanoscopic" parameterization (Type II parameterization [Urbina-Villalba, 2015; 2016; Rahn-Chique, 2017b]) overestimates the value of the aggregation rate, but reproduces the qualitative behavior of the systems up to 500 mM NaCl. Instead, the implementation of "macroscopic" (Gibbs) isotherms (Type I parameterization) along with the assumption of deformable droplets replicates the order of magnitude of the rates but predicts non-flocculating emulsions between 300 and 900 mM NaCl. This fact is incompatible with the experimental variation of the absorbance.

In the case of d/w emulsions, the rates of two-particle calculations using a T-II parameterization increase monotonically with the ionic strength. They show the same order of magnitude of experimental rates, but fail to duplicate the decrease observed above 500 mM NaCl. These differences suggest that other processes might be occurring along with the flocculation of the drops.

However, the procedure followed to establish T-II parameterization was devised back in 2014 in order to justify the appearance of a maximum surface potential in h/w systems. It takes into account that surfactant adsorption increases as a function of the ionic strength until maximum surface excess, and then decreases due to the screening of the resulting double layer.

Among other initiatives, we found convenient to confirm the existence of a maximum of surface potential in dodecane-in-water nanoemulsions. Since the maximum is expected between 20 – 50 mM NaCl, the corresponding aggregation rates cannot be evaluated experimentally. Therefore, the change in the average radius of the emulsions as a function of time was used to assess the stability of these systems.

## 2. EXPERIMENTAL PROCEDURE

Dodecane (Merck, 98%) was purified using an alumina column. SDS, sodium chloride (Merck, 99.5%) and iso-pentanol (Scharlau Chemie, 99%) were used as received. Distilled water was deionized using a Millipore Simplicity apparatus.

Nanoemulsions were prepared using a phase inversion composition method [Rahn-Chique, 2012]. A mixture of liquid crystal solution and oil ($\phi$ = 0.84, [SDS] = 10 wt%, [NaCl] = 8% wt%, and [isopentanol] = 6,5 wt%) previously pre-equilibrated, was suddenly diluted until $\phi$ = 0.44, to obtain 120-150 nm drops of oil. This mother nanoemulsion was further diluted with pure water until $\phi$ = 0.02, and with an appropriate brine/surfactant solution until $\phi$ = 3.2 x 10$^{-4}$.

The zeta potential of the drops corresponding to 22 systems at [SDS] = 7.5 mM, and 0 < [NaCl] < 900 mM was measured using a Zetasizer Nano from Malvern. Additional measurements at 0.5 mM SDS were also made for the lowest 16 salinities.

The initial radius of the emulsions without salt was determined employing a Beckman-Coulter LS-230 (blanks). Following, the appropriate aliquot of brine was added, and the radius appraised again. The variation of the average radius of the emulsion as a function of time was then followed using a BI-200SM goniometer from Brookhaven Instruments.

## 3. RESULTS AND DISCUSSION

Fig. 1 compares the surface potentials obtained for dodecane (at 0.5 and 7.5 mM SDS) with the ones previously measured for hexadecane using a Delsa 440 SX (Coulter). The newer Zetasizer allows to reach higher ionic strengths ( > 200 mM NaCl). The curves have similar shapes showing a maximum between 20 and 50 mM NaCl. The data of hexadecane, differs appreciably for the two surfactant concentrations tested (0.5 and 7.5 mM SDS), while the ones of d/w emulsions appear to correspond to the same body of data, suggesting a minor concentration dependence.

Figs 2 and 3 compare the experimental surface potentials obtained for dodecane, with the predictions of the "macroscopic" (Type I) and "nanoscopic" (Type II) adsorption isotherms. Unlike their h/w analogs, the values of the

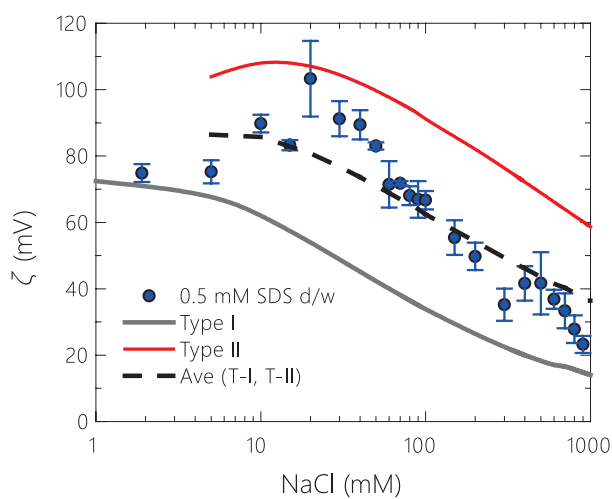

Figure 2. Comparison between the surface potential of 0.5 mM SDS d/w emulsions, and the predictions of Type I and Type II isotherms.

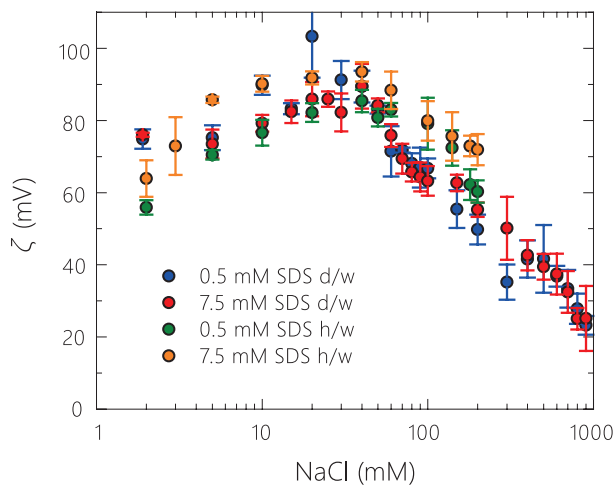

Figure 1. Dependence of the surface potential of d/w and h/w nanoemulsions at 7.5 y 0.5 mM SDS on the ionic strength.

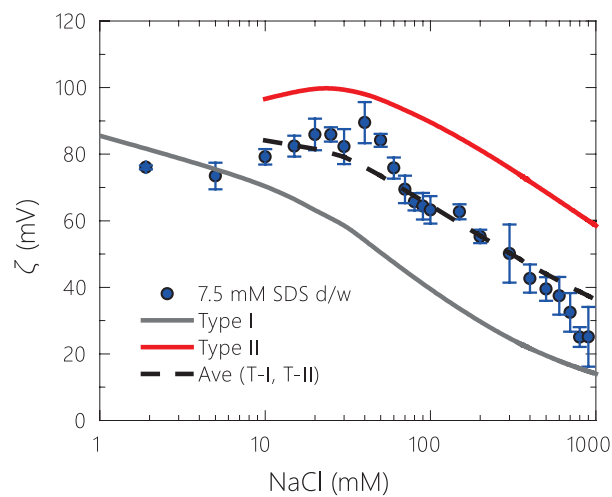

Figure 3. Comparison between the surface potential of 7.5 mM SDS d/w emulsions, and the predictions of Type I and Type II isotherms.

Daniela Díaz, Kareem Rahn-Chique, German Urbina-Villalba

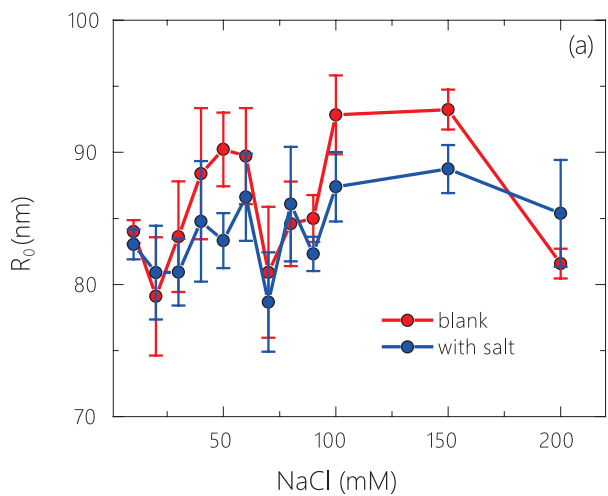

Figure 4(a). Initial average radius ($R_0$) of d/w nanoemulsions before and after the addition of salt. (a) 0.5 mM SDS.

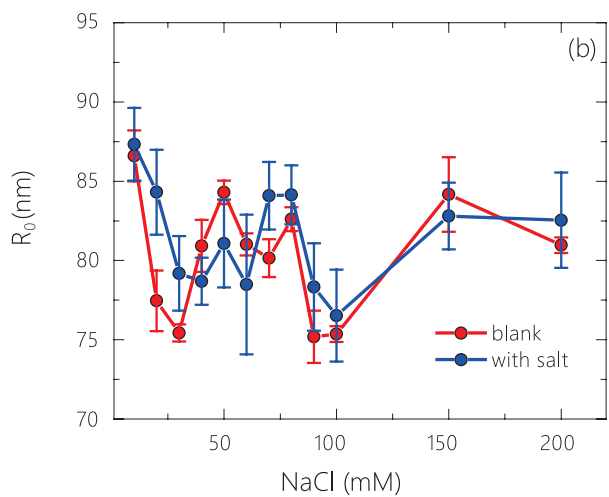

Figure 4(b). Initial average radius ($R_0$) of d/w nanoemulsions before and after the addition of salt. (b) 7.5 mM SDS.

surface potentials corresponding to d/w nanoemulsions lay between both models. Average values calculated with the forecasts of the models are illustrated in these figures with a dash curve, which nicely coincides with the experimental data. There is a slight tendency to favor Type I potential at very high or very low salt concentrations, and Type II potential at intermediate ionic strengths.

Since the amount of salt used in the evaluation of stability was very low, it was expected that the variation of the average radius of the emulsions could be very small during a 5 minutes period. Hence, it was necessary to confirm that the addition of salt did not alter significantly the initial average radius. For that purpose, the radius of the synthesized nanoemulsions was measured before and after the addition of salt. Figures 4(a) and 4(b) illustrate the corresponding data for 0.5 and 7.5 mM SDS. Fortunately, low concentrations of sodium chloride do not alter the average radius of the emulsions appreciably, though variations as high as 5 nm are observed. The greatest differences occur at 40, 50, 60, 100, and 150 mM NaCl for 0.5 mM SDS, and 20, 30, 50 and 60 mM NaCl in the case of 7.5 mM SDS.

Figures 5(a) and 5(b) show typical changes in the diameter of the emulsions as a function of time for very low ionic strength. These figures correspond to the systems with lowest (5(a)) and highest (5(b)) standard deviations of the 7.5 mM SDS set. The size does not change monotonically as it occurs at higher salinities ([NaCl] > 300 mM) where aggregation prevails. Fig. 5(c) illustrates this alternative situation for 400 mM NaCl. In the latter case the increment of the radius is monotonous and significant, as reflected by much larger correlation coefficients.

According to the change in absorbance vs. time for 100 and 200 mM NaCl [Diaz, 2018], micelle solubilization predominates during at least 200 s in the case of 7.5 mM SDS d/w nanoemulsions. However, as shown by Fig. 5(b), the slope of R vs. t can increase, decrease or stay the same. In fact, 7 out of 12 salinities show a positive slope in the case of 0.5 mM SDS, and 8 out of 12 in the case of 7.5 mM SDS.

As observed in Figs 5(a) and 5 (b), the variation of the average radius appears to be random, and possibly connected to the inherent limitations of the light scattering technique. Moreover, correlation coefficients are very deficient, and the amount of data insufficient to establish a conclusion of statistical significance. However, it is noteworthy that the average radius increases at certain times. This can be the result of the simultaneous occurrence of Ostwald ripening and solubilization. According to previous calculations on the subject [Urbina-Villalba, 2009], the average radius produced by the Ostwald ripening process should increase ***exclusively*** when the number of drops decreases as a consequence of coalescence and/or complete solubilization of the drops, but ***not during the exchange of molecules*** between the drops. In other words, the exchange of molecules typical of the ripening process leads to a ***decrease*** of the average radius.

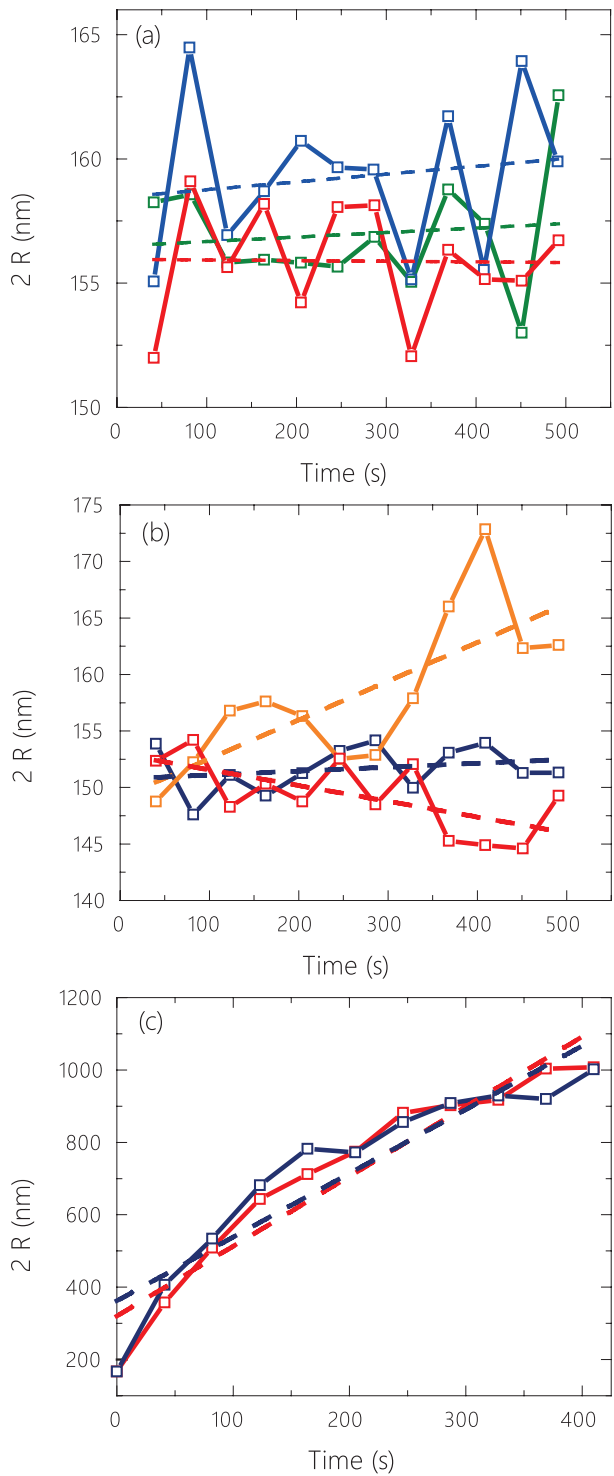

Figure 5. Change in the average diameter of three d/w nanoemulsions stabilized with 7.5 mM SDS. (a) 40 mM NaCl. Regression coefficients ($r^2$) of 0.021, 0.0124, and 0.0003 are obtained. (b) 100 mM NaCl. Regression coefficients ($r^2$) of: 0.56, 0.06, and 0.42 are found. (c) 400 mM NaCl. Regression coefficients ($r^2$) of 0.86 and 0.91 are observed.

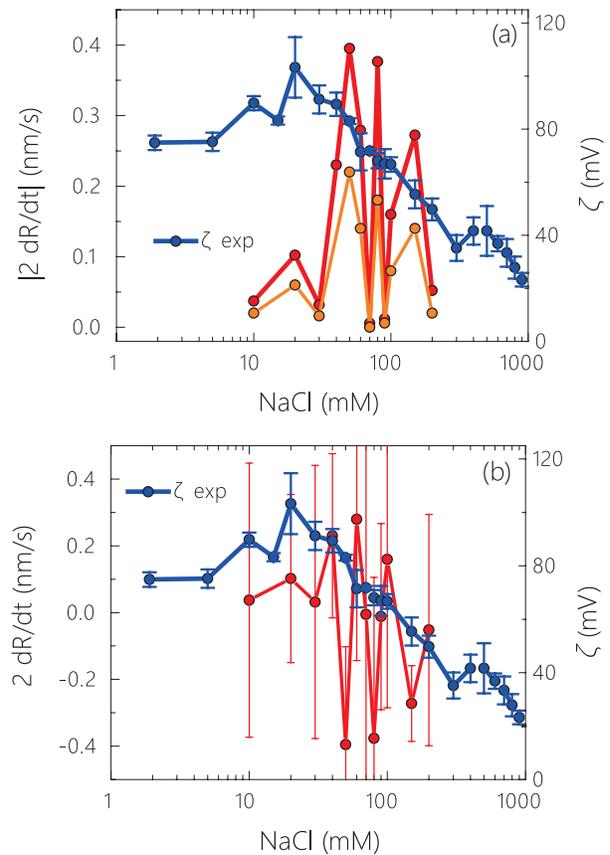

Figure 6. Comparison between the zeta potential of 0.5-mM SDS d/w nanoemulsions (blue curve) and the change in the average diameter (2 x Radius) of the resulting emulsions.
(a) The red curve illustrates the absolute value of the average slope of three measurements. The orange curve results from using the data of the three measurements to obtain one average slope. The values of the average slope were multiplied by 25 for better observation (orange curve).
(b) Change in the average slope of the three curves with sign and corresponding standard deviation.

During solubilization, the exchange of molecules occurs between drops and micelles. Therefore, the average radius of the drops ***must necessarily decrease***. Nevertheless, it might temporarily increase when the smallest drops finally disappear as in the case of ripening. Hence, dR/dt should markedly depend on the number of drops, and might be positive, negative or even zero.

In order to have a quantitative measurement of stability, the absolute value of the average slope of three measurements was computed despite their extremely poor correlation coefficients (red curve in Figs. 6(a) and 7(a)). Additionally, the linear regression of the whole body of data was also computed (orange curve in Figs. 6(a) and 7(a)). Fig-

Daniela Díaz, Kareem Rahn-Chique, German Urbina-Villalba

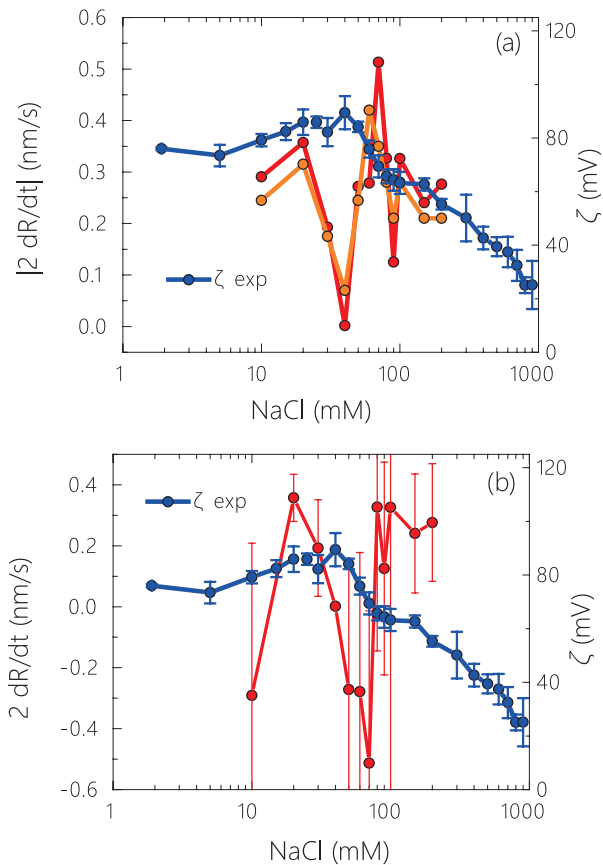

Figure 7. Comparison between the zeta potential of 7.5-mM SDS d/w nanoemulsions (blue curve) and the change in the average diameter ( 2 x Radius) of the resulting emulsions.
(a) The red curve illustrates the absolute value of the average slope of three measurements. The orange curve results from using the data of the three measurements to obtain one average slope. The values of the average slope were multiplied by 35 for better observation (orange curve).
(b) Change in the average slope of the three curves with sign and corresponding standard deviation.

ure 6 corresponds to 0.5 mM SDS. It is clear that there does not appear to be any correlation between the maximum of the surface potential and the change of the average radius. In fact, when the sign of the average slope is considered, it is possible to draw a straight line through all the errors bars (Fig. 6(b)). The situation does not change if all the experimental points are used to compute one average slope for all concentrations. Conversely, in the case of 7.5 mM SDS the minimum slope and the maximum surface potential coincides at 40 mM NaCl (Figs 7(a) and 7(b)). The average radius appears to decrease between 5 and 70 mM NaCl, and stay the same from 100 to 200 mM NaCl.

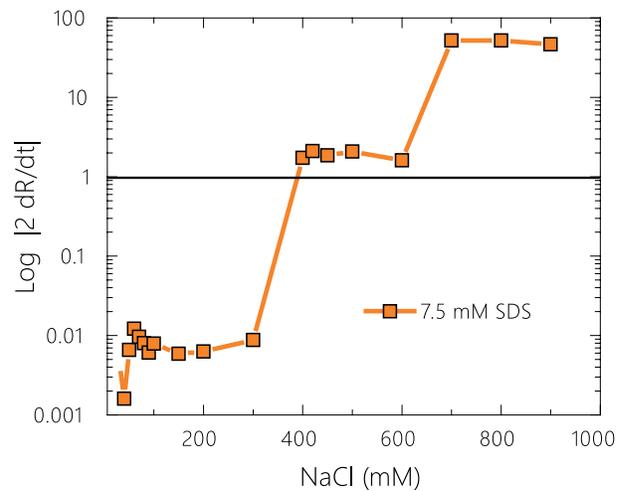

Figure 8. Order of magnitude variation of the average diameter of 7.5 mM-SDS d/w emulsions as a function of the NaCl concentration. Three regimes are clearly observed.

A very insightful graph results from plotting the absolute value of the average slope vs. the salt concentration in a log-linear scale (Figs. 8). Three plateaus appear. They illustrate the relative stability of the systems subject to the regimes of (1) solubilization/ripening, (2) flocculation/coalescence, and (3) surfactant crystallization.

### 4. CONCLUSIONS

The surface potential of dodecane drops in d/w nanoemulsions shows an intermediate behavior between the prediction of macroscopic isotherms and the forecast of the formalism previously developed for surfactant adsorption on nanoemulsion drops.

The electric surface potential of d/w nanoemulsions shows a maximum value around 20 and 40 mM NaCl for 0.5 and 7.5 mM SDS, respectively.

For a surfactant concentration of 7.5 mM SDS, there appears to be a relationship between the overall stability of the nanoemulsions and the maximum electric surface potential found. This phenomenon is not related to the flocculation of the drops, and occurs in conditions where micelle solubilization prevails.